# THE ONEFOLD TRUTH

Rainer Burghardt[*]

Contents

Some features of the Schwarzschild and Kruskal metric are being discussed under the assumption that the Schwarzschild model can be explained geometrically.

---

[*] e-mail: arg@onemail.at, home page: http://arg.at.tf/

# 1. INTRODUCTION

In former papers [1,2] we have shown that the Schwarzschild model can be embedded in a flat five-dimensional space by the use of two correlated surfaces, generated by the rotation of Schwarzschild's and Neil's parabola. This double surface can be derived from a single surface by projection techniques. These two surfaces have a common Riemann and Ricci tensor. If we cut off all we do not need for the four-dimensional description we obtain the physical surface. The physical surface is the place of all possible observations and measurements. We are not able to experience the extra dimensions, but it might be a good mathematical method to use a higher-dimensional theory and embedding techniques to describe four-dimensional physics. In this paper we make use of this strategy. We analyze the Schwarzschild metric in the standard and Kruskal forms and we make four assumptions:

(i) We interpret the Schwarzschild metric geometrically as the common metric of two correlated surfaces embedded in a five-dimensional flat space.

(ii) Coordinate transformations are arbitrary and a matter of convenience. They do not alter the invariant geometrical structure of the theory.

(iii) We will use vectors as field strengths representing the normal and odd curvatures of these surfaces.

(iv) Any transitions to new states of motion are performed by local Lorentz transformations.

The consequence of these assumptions is that the Schwarzschild model will not supply Black Hole physics and the Kruskal metric will not provide interior solutions.

# 2. THE SCHWARZSCHILD STANDARD METRIC

The Schwarzschild metric can be written as

$$ds^2 = \rho_i \rho_k d\varepsilon^i d\varepsilon^k, \quad i = k \tag{2.1}$$

with

$$\rho_1 = \rho = \sqrt{\frac{2r^3}{M}}, \quad \rho_2 = r, \quad \rho_3 = r\sin\vartheta, \quad \rho_4 = \rho\cos\varepsilon$$
$$\varepsilon^1 = \varepsilon, \quad \varepsilon^2 = \vartheta, \quad \varepsilon^3 = \varphi, \quad \varepsilon^4 = i\psi \tag{2.2}$$

$\rho$ is the curvature vector and $\varepsilon$ the angle of ascent of the Schwarzschild parabola.

Flamm's paraboloid is generated by rotating the Schwarzschild parabola about the directrix through the angles $\vartheta$ and $\varphi$. Another piece of the four-dimensional surface is generated by an imaginary rotation about the r-axis, i. e., the symmetry axis of the Schwarzschild paraboloid. Neil's paraboloid is generated by a similar procedure. From (2.1) we obtain

$$ds^2 = \rho^2 d\varepsilon^2 + r^2 d\vartheta^2 + r^2 \sin^2\varepsilon\, d\varphi^2 + \rho^2 \cos^2\varepsilon\, d\psi^2 . \tag{2.3}$$

$\varepsilon$ is taken to be cw and has the range [π/2, 0].

$$\sin\varepsilon = -\sqrt{\frac{2M}{r}} \tag{2.4}$$

equals the velocity of a freely falling observer. Differentiating this expression, we get

$$\rho d\varepsilon = \frac{1}{\cos\varepsilon} dr = \frac{1}{\sqrt{1-2M/r}} dr . \tag{2.5}$$

Replacing $\rho d\psi$ in (2.3) with the Schwarzschild co-ordinate time dt we obtain the Schwarzschild metric in the standard notation. Since the Schwarzschild parabola has the vertex at r = 2M ($\rho$ = 4M, $\varepsilon$ = π/2) no geometry is defined for r < 2M. The Schwarzschild metric has a boundary at r = 2M as a consequence of assumption (i). The Schwarzschild metric is regular everywhere on the physical surface, also at r = 2M. From the equation of the Schwarzschild parabola

$$R^2 = 8M(r - 2M) , \tag{2.6}$$

R being the co-ordinate of the extra dimension of the flat embedding space, we get instead of (2.5)

$$\rho d\varepsilon = -\frac{1}{\sin\varepsilon} dR = \sqrt{1 + \frac{R^2}{16M^2}} dR , \tag{2.7}$$

which shows that the tangent vector of the physical surface at the boundary of the geometry is simply dR and is normal to the symmetry axis of the Schwarzschild parabola. The 'singularity' at 2M can be removed by a suitable choice of the parameters. This was pointed out by Einstein and Rosen [3]. They defined a new variable, which differs from our R by the factor $\sqrt{8M}$. The two solutions of (2.6) are the two sheets of the same space, connected by the bridge at R = 0. The second sheet corresponds to the negative branch of the Schwarzschild parabola. Since the sign of the ascent of this branch is opposite to the sign of the ascent of the upper branch the velocity of a free object would be positive and the action of the central mass repulsive. Since antigravitation is not known to occur in nature we exclude the second branch from the theory. The complete Schwarzschild model consists of the exterior and interior solution. The latter has to be matched to the exterior solution in such a way that the pressure of the source does not grow infinitely large [4]. This could be achieved by adjusting the aperture angle of the cap of sphere which represents the space-like part of the physical surface in a convenient manner. The complete solution is free from 'singularities' and exhibits no peculiarities. It is a good model for approximately describing stellar objects. The interior solution also can be embedded in a five-dimensional flat space [2]. It turns out that the energy-momentum tensor consists of the generalized second fundamental forms of the embedded surfaces and thus the matter can be interpreted as fields. The generalized Codazzi equations of the surfaces are the field equations for these matter fields.

The purist geometrical interpretation of the Schwarzschild metric demands that no incoming particle can cross the boundary of the geometry. It is widely accepted that any particle incoming from infinity approaches the speed of light at r = 2M. But for particles starting from a position different from infinity and falling to the center of gravitation several points of view are known from literature. In most of these concepts [5 -20] the radial co-ordinate r may be interpreted as the path the observers can travel on, although r and t interchange their meaning for r < 2M. There was a discussion if particles can cross the Schwarzschild radius with a velocity lower than the speed of light by Janis, Cavalleri and Spinelli [12], Tereno [21] and Crawford [22]. Mitra [23-27] showed that the velocity of a particle approaches the speed of light at r = 2M independently of the position it was released from. This view is supported by the following consideration: A particle in-falling from infinity has the velocity $v_0 = -\sqrt{2M/r_0}$ at the radial position $r_0$. Another particle is released from this position at the moment the first particle is passing. The difference of the velocity of these particles at this moment is $v_0$. Both particles are exposed to the same gravitational force $E = -M/\left(r^2\sqrt{1-2M/r}\right)$ on their travel to the center of gravitation. The difference of the velocities decreases during the motion due to Einstein's composition law of velocities. For particles released from different positions we get as their velocities

$$v(r, r_0) = \frac{-\sqrt{\frac{2M}{r}} - \left(-\sqrt{\frac{2M}{r_0}}\right)}{1 - \sqrt{\frac{2M}{r}}\sqrt{\frac{2M}{r_0}}} \quad , \tag{2.8}$$

where $v(2M, r_0) = -1$, $v(r_0, r_0) = 0$. Fig. 1 shows some examples. Moreover, Mitra has shown in [28] that the Schwarzschild radius for Black Holes is r = 0 and the mass of a Black Hole has to be M = 0.

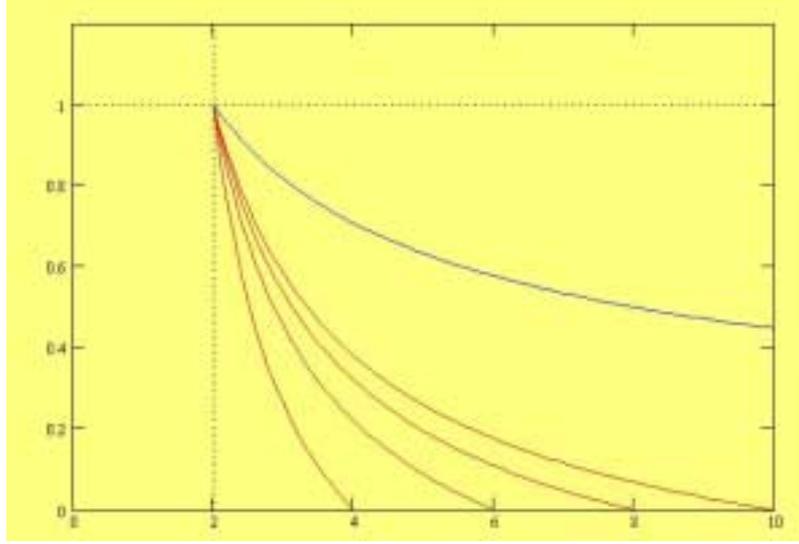

**Fig. 1**

# 3. THE KRUSKAL METRIC

Kruskal [29] has found a co-ordinate system that is regular at r = 2M. The Schwarzschild metric in this co-ordinate system can be written as

$$ds^2 = \gamma^2 \left(du^{1^2} + du^{4^2}\right) + r^2 d\vartheta^2 + r^2 \sin^2\vartheta\, d\varphi^2 \,, \tag{3.1}$$

$$\gamma^2 = \frac{32 M^3}{r} e^{-\frac{r}{2M}} \,. \tag{3.2}$$

This co-ordinate system covers four sectors with

$$\text{I } \begin{aligned} u^1 &= Y\cos i\chi \\ u^4 &= Y\sin i\chi \end{aligned}, \quad \text{II } \begin{aligned} u^1 &= Y\sin i\chi \\ u^4 &= -Y\cos i\chi \end{aligned}, \quad \text{III } \begin{aligned} u^1 &= -Y\cos i\chi \\ u^4 &= -Y\sin i\chi \end{aligned}, \quad \text{IV } \begin{aligned} u^1 &= -Y\sin i\chi \\ u^4 &= Y\cos i\chi \end{aligned}, \tag{3.3}$$

where

$$Y(r) = \frac{\sqrt{1-2M/r}}{\sqrt{2M/r}} e^{\frac{r}{4M}}, \quad Y(2M) = 0, \quad \chi = \frac{t}{4M}. \tag{3.4}$$

The imaginary circle $u^{1^2} + u^{4^2} = Y^2$ consists of four branches of hyperbolae of constant curvature and two null lines for r = 2M. Plainly planely drawn one obtains the Kruskal diagram. Sectors I and III are said to describe the physics of an incoming and outgoing rocket and II and IV the interior region of the Schwarzschild metric. As we do not believe that a co-ordinate transformation changes the geometrical content of the theory we have to search for another explanation for the Kruskal metric. Evidently, the connexion coefficients derived from (2.3) have an invariant geometrical meaning. The

$$A_{mn}{}^s = B_{mn}{}^s + C_{mn}{}^s + E_{mn}{}^s$$
$$B_{mn}{}^s = b_m B_n b^s - b_m b_n B^s, \quad C_{mn}{}^s = c_m C_n c^s - c_m c_n C^s, \quad E_{mn}{}^s = u_m u_n E^s - u_m E_n u^s, \qquad (3.5)$$
$$m_n = \{1,0,0,0\}, \quad b_m = \{0,1,0,0\}, \quad c_m = \{0,0,1,0\}, \quad u_m = \{0,0,0,1\}$$

represent the curvatures of the physical surface:

$$B_m = \left\{\frac{1}{\rho_2}\cos\varepsilon, 0, 0, 0\right\}, \quad C_m = \left\{\frac{1}{\rho_3}\sin\vartheta\cos\varepsilon, \frac{1}{\rho_3}\cos\vartheta, 0, 0\right\}, \quad E_m = \left\{\frac{1}{\rho_4}\sin\varepsilon, 0, 0, 0\right\}. \qquad (3.6)$$

By the use of an extra dimension these expressions can be complemented with

$$M_0 = A_{11} = \frac{1}{\rho_1}, \quad B_0 = A_{22} = \frac{1}{\rho_2}\sin\varepsilon, \quad C_0 = \frac{1}{\rho_3}\sin\vartheta\sin\varepsilon, \quad E_0 = A_{44} = -\frac{1}{\rho_4}\cos\varepsilon. \qquad (3.7)$$

In agreement with assumption (iii) the $A_{mn}$ are the generalized second fundamental forms of the physical surface. A co-ordinate transformation does not have an effect on this structure but a Lorentz transformation may be correlated to this co-ordinate transformation. By differentiation of (3.3) for sector I we obtain

$$du^1 = \cos i\chi \, dY - Y\sin i\chi \, di\chi, \quad du^4 = \sin i\chi \, dY + Y\cos i\chi \, di\chi. \qquad (3.8)$$

Evaluating dY with (3.4) and multiplying with $\gamma$ we get the rotated vectors

$$dx^{1'} = \cos i\chi \, dx^1 - \sin i\chi \, dx^4, \quad dx^{4'} = \sin i\chi \, dx^1 + \cos i\chi \, dx^4, \qquad (3.9)$$

from which we can read off the components of the Lorentz transformation

$$L_1^{1'} = \cos i\chi, \quad L_4^{1'} = -\sin i\chi, \quad L_1^{4'} = \sin i\chi, \quad L_4^{4'} = \cos i\chi. \qquad (3.10)$$

Thus the velocity of the outgoing Kruskal rocket is

$$v_K = \text{th}\chi. \qquad (3.11)$$

In (3.9) we have used the Schwarzschild standard expressions for the radial line element and the local time interval

$$dx^1 = \frac{1}{\cos\varepsilon}dr, \quad dx^4 = \cos\varepsilon\, idt, \quad \cos\varepsilon = \sqrt{1 - 2M/r} \ . \tag{3.12}$$

Calculating $dx^{1'}/dx^{4'}$ with (3.9) we obtain by using the proper times $dx^{4'} = id\tau'$, $dx^4 = id\tau$ Einstein's composition law for velocities

$$\frac{dx^{1'}}{d\tau'} = \frac{v_R + v_K}{1 + v_R v_K} \ , \tag{3.13}$$

where $v_R = dx^1/d\tau$ is the unspecified velocity of an observer in radial motion. For time reversion $\chi \to -\chi$ we obtain the velocity of an incoming rocket $v_K = -\text{th}\chi$ and an analogous composition law. For sectors II and IV we get an equivalent relation to (3.13) with the velocities

$$v_K = \text{cth}\chi \ . \tag{3.14}$$

The motion is tachyonic [17]. The rocket is at rest at r = 2M with infinite speed and it accelerates to the speed of light in approaching infinity. For time reversion the rocket is at rest with the speed of light at infinity. On its way to the center of gravitation it will slow down to infinite speed. Fig. 2 is a diagram for both kinds of velocities (3.11) and (3.14).

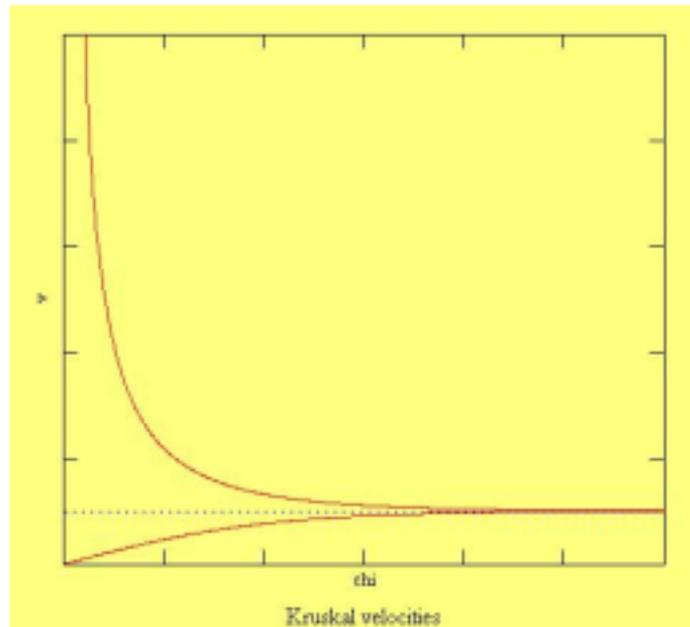

**Fig. 2**

All these considerations refer to the tangent space of the exterior solution. Excluding the unphysical tachyonic sectors, Equs. (3.11) and (3.13) show that a Kruskal rocket is not able to cross the boundary of the geometry. The velocity of an incoming rocket tends to the speed of light in approaching r = 2M.

# 4. THE KRUSKAL ACCELERATION

In the preceding chapter we have shown that the transition from the standard form of the Schwarzschild metric to the Kruskal form can be interpreted as Lorentz transformation with non-constant velocity parameter acting in the tangent space of the Schwarzschild geometry. This results in an accelerated motion of observers. In the following, we will derive forces, which are responsible for this acceleration. Firstly, we remark that the curvature properties of the geometry are invariant under the Lorentz transformation (3.10). In the new reference system the curvatures (3.6) get time-like components

$$B_m = \left\{ \frac{1}{\rho_2} \cos\varepsilon \cos i\chi, 0, 0, \frac{1}{\rho_2} \cos\varepsilon \sin i\chi \right\}$$

$$C_m = \left\{ \frac{1}{\rho_3} \sin\vartheta \cos\varepsilon \cos i\chi, \frac{1}{\rho_3} \cos\vartheta, 0, \frac{1}{\rho_3} \sin\vartheta \cos\varepsilon \sin i\chi \right\} . \qquad (4.1)$$

$$E_m = \left\{ \frac{1}{\rho_4} \sin\varepsilon \cos i\chi, 0, 0, \frac{1}{\rho_4} \sin\varepsilon \sin i\chi \right\}$$

They transform as vectors under (3.10). Since the connexion coefficients are constructed with these quantities and also with the tetrads (3.5), the connexion coefficients transform as three-rank tensors:

$$A_{n'm'}{}^{s'} = L_{n'm's}^{nms'} A_{nm}{}^{s} . \qquad (4.2)$$

The covariant derivative transforms as

$$\Phi_{m'||n'} = L_{m'n'}^{mn} \Phi_{m||n} = \left[ \Phi_{m'|n'} - L_s^{s'} L_{m'|n'}^{s} \Phi_{s'} \right] - A_{n'm'}{}^{s'} \Phi_{s'} . \qquad (4.3)$$

With the definition

$$\Phi_{m'\underset{1}{||}n'} = \Phi_{m'|n'} - L_{n'm'}{}^{s'} \Phi_{s'}, \quad L_{n'm'}{}^{s'} = L_s^{s'} L_{m'|n'}^{s} \qquad (4.4)$$

and the set of graded derivatives

$$m_{m'||n'} = m_{m'|n'} - L_{n'm'}{}^{s'} m_{s'} = 0$$
$$\underset{1}{}$$

$$b_{m'||n'} = b_{m'|n'} - L_{n'm'}{}^{s'} b_{s'} = 0$$
$$\underset{2}{}$$

$$c_{m'||n'} = c_{m'|n'} - \left(L_{n'm'}{}^{s'} + B_{n'm'}{}^{s'}\right) c_{s'} = 0$$
$$\underset{3}{}$$

$$u_{m'||n'} = u_{m'|n'} - \left(L_{n'm'}{}^{s'} + B_{n'm'}{}^{s'} + C_{n'm'}{}^{s'}\right) u_{s'} = u_{m'|n'} - L_{n'm'}{}^{s'} u_{s'} = 0$$
$$\underset{4}{}$$

(4.5)

we are able to simplify the calculations considerably. (4.5) shows that the tetrads are parallel transported with respect to the graded transport law. The unit vectors of the 1- and 4-direction of the static system measured by the Kruskal observers have the components

$$m_{n'} = \{\cos i\chi, 0, 0, \sin i\chi\} \quad u_{m'} = \{-\sin i\chi, 0, 0, \cos i\chi\}. \tag{4.6}$$

Their own unit vectors are

$$'m_{n'} = \{1, 0, 0, 0\}, \quad 'u_{m'} = \{0, 0, 0, 1\} \tag{4.7}$$

We calculate the Lorentz term by using (4.6) or (4.7)

$$L_{m'n'}{}^{s'} = h_{m'n'} K^{s'} - h_{m'}{}^{s'} K_{n'}, \quad h_{m'n'} = 'm_{m'} 'm_{n'} + 'u_{m'} 'u_{n'} = m_{m'} m_{n'} + u_{m'} u_{n'}, \tag{4.8}$$

wherein the Kruskal acceleration is

$$K_{n'} = \frac{1}{4M\cos\varepsilon} m_{n'} = \frac{1}{4M\cos\varepsilon} \{\cos i\chi, 0, 0, \sin i\chi\} \tag{4.9}$$

and measured in the static system

$$K_n = \frac{1}{4M\cos\varepsilon} m_n = \frac{1}{4M\cos\varepsilon} \{1, 0, 0, 0\}. \tag{4.10}$$

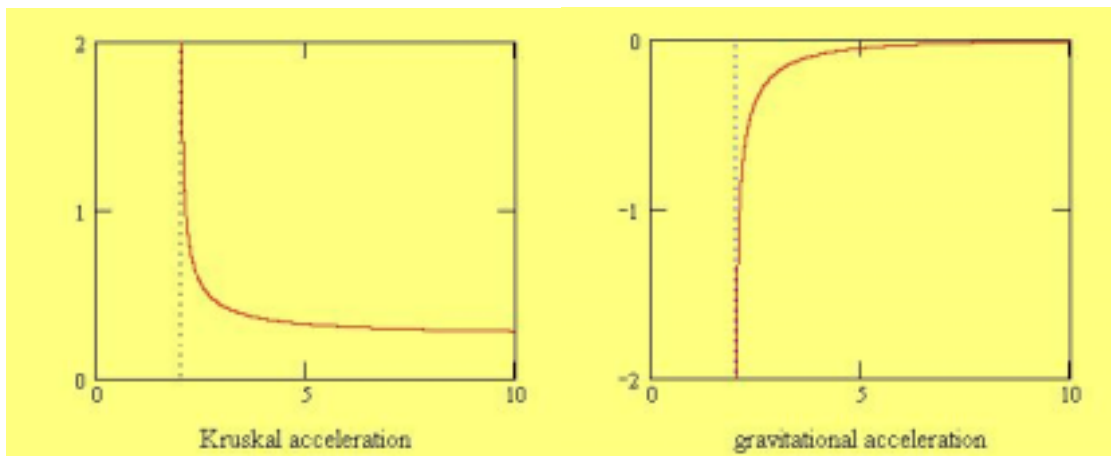

Kruskal acceleration      gravitational acceleration

**Fig. 3**

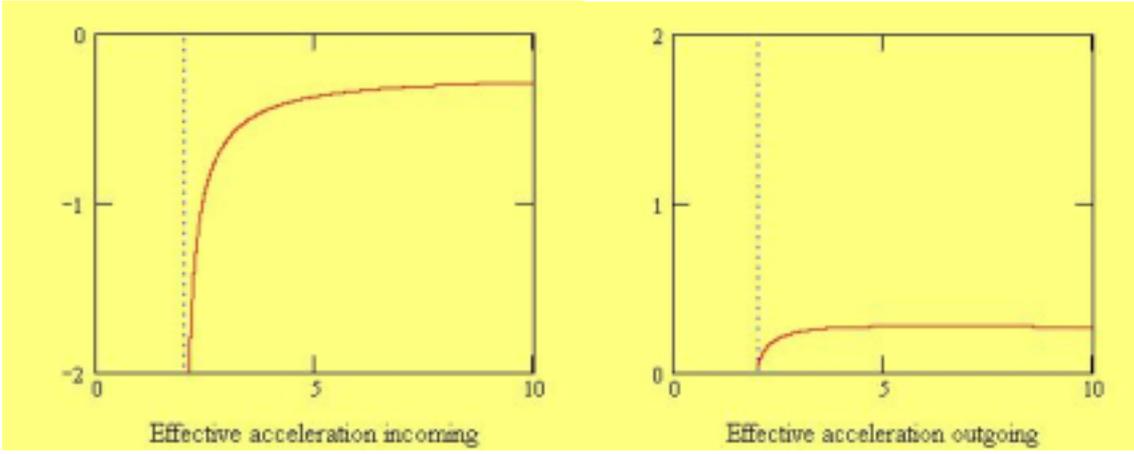

**Fig. 4**

In Fig. 3 the Kruskal and the gravitational accelerations (3.6) are plotted, in Fig. 4 the effective accelerations for an incoming and outgoing rocket taking into account the gravitational acceleration. Evidently we have

$$'u_{\alpha'\|n'}\,'u^{n'} = -\left[A_{4'\alpha'}{}^{4'} + L_{4'\alpha'}{}^{4'}\right] = E_{\alpha'} + K_{\alpha'}, \quad \alpha = 1,2,3 \qquad (4.11)$$

and the Kruskal acceleration seems to be an economical acceleration for rockets.

Since the Ricci tensor is invariant under the Lorentz transformation $R_{m'n'} = L_{m'n'}^{m\,n} R_{mn} = 0$ we get with

$$R_{m'n'} = A_{m'n'}{}^{s'}{}_{\|s'} - A_{n'\|m'}{}_1 - A_{r'm'}{}^{s'} A_{s'n'}{}^{r'} + A_{m'n'}{}^{s'} A_{s'} \qquad (4.12)$$

the same equations that we derived for the static Schwarzschild system in former papers:

$$R_{n'm'} = -\left[B_{m'\|n'}{}_2 + B_m B_{n'}\right] - b_{m'} b_{n'}\left[B^{r'}{}_{\|r'}{}_2 + B^{r'}B_{r'}\right] - \left[C_{m'\|n'}{}_3 + C_m C_{n'}\right] - c_{m'} c_{n'}\left[C^{r'}{}_{\|r'}{}_3 + C^{r'}C_{r'}\right] +$$
$$+ \left[E_{m'\|n'}{}_4 - E_m E_{n'}\right] + u_{m'} u_{n'}\left[E^{r'}{}_{\|r'}{}_4 - E^{r'}E_{r'}\right] = 0, \quad B_{[m'\|n']}{}_2 = 0, \quad C_{[m'\|n']}{}_3 = 0, \quad E_{[m'\|n']}{}_4 = 0 \qquad (4.13)$$

These equations stand for the prediction that an accelerated observer makes for the physics of static observers. The components of the curvatures are expressed in the reference system of the accelerated observers. One should obtain the same Ricci tensor by an inhomogeneous transformation of the connexion coefficients

$$R_{m'n'} = 'A_{m'n'}{}^{s'}{}_{|s'} - 'A_{n'|m'} - 'A_{r'm'}{}^{s'}\,'A_{s'n'}{}^{r'} + 'A_{m'n'}{}^{s'}\,'A_{s'}$$
$$'A_{n'm'}{}^{s'} = L_{n'm's}^{nm\,s'} A_{nm}{}^{s} + L_{n'm'}{}^{s'} \qquad (4.14)$$

This implies the condition

$$L_{m'n'\ |s'}^{\ \ s'} - L_{s'n'\ |m'}^{\ \ s'} - L_{r'm'}^{\ \ s'}L_{s'n'}^{\ \ r'} + L_{m'n'}^{\ \ s'}L_{r's'}^{\ \ r'} + 2A_{[m'r']}^{\ \ \ s'}L_{s'n'}^{\ \ r'} = 0 \ . \qquad (4.15)$$

Inserting (4.8) into the above equation we obtain the field equation for the Kruskal acceleration

$$K^{s'}_{\ \underset{1}{\|}s'} - K^{s}E_{s'} = 0 \ . \qquad (4.16)$$

This is just the subequation of the field equations if the connexion coefficients have been transformed inhomogeneously. In the five-dimensional representation (a = 0,1,…,4) all subequations of the field equations decouple [1]:

$$B_{b'\underset{2}{\|\|}a'} + B_{b'}B_{a'} = 0, \quad B^{c'}_{\ \underset{2}{\|\|}c'} + B^{c'}B_{c'} = 0$$

$$C_{b\underset{3}{\|\|}a} + C_b C_a = 0, \quad C^{c'}_{\ \underset{3}{\|\|}c'} + C^{c'}C_{c'} = 0 \ . \qquad (4.17)$$

$$E^{c'}_{\ \underset{1}{\|\|}c'} - E^{c'}E_{c'} = 0, \quad K^{c'}_{\ \underset{1}{\|\|}c'} - K^{c'}E_{c'} = 0$$

The equations for the gravitational acceleration and the Kruskal acceleration are expressed in a similar way, but one has to bear in mind that the Kruskal acceleration is due to a structure *on* the physical surface while the other equations describe the structure *of* the surface.

## 5. SUMMARY

In this paper we have shown that the velocity of observers, freely falling or Kruskal accelerated to the center of gravitation, tends to the speed of light for r → 2M. No object is able to cross the event horizon of the Schwarzschild geometry. This agrees with our assumptions made at the beginning of the paper. The Schwarzschild radius fixes the boundary of the geometry. For r < 2M no geometry is defined by the exterior Schwarzschild solution, independently of the co-ordinate system in use. Interpreting the Schwarzschild metric strictly geometrically no Black Hole physics can be derived from the Schwarzschild solution.

I am indebted to Prof. H.-J. Treder for his kind interest in this paper.